\documentstyle[12pt]{article}

\title{Supersymmetric approach  
for generating quasi-exactly solvable potentials
with arbitrary two known eigenstates}
      
\author{V. M. Tkachuk \\
Ivan Franko Lviv National University, Chair of Theoretical Physics \\ 
		 12 Drahomanov Str., Lviv UA--79005, Ukraine\\
		 E-mail: tkachuk@ktf.franko.lviv.ua }

\begin{document}

\maketitle

\begin{abstract}
Using supersymmetric quantum mechanics we construct 
the quasi-exactly solvable (QES) potentials with 
arbitrary two known eigenstates.
The QES potential and the wave functions of the two energy levels are
expressed by some generating function
the properties of which determine the state numbers of these levels.
Choosing different generating functions we present a few explicit examples
of the QES potentials.

\end{abstract}

PACS number(s): 03.65.Ge; 11.30.Pb

\section{Introduction}
From the earlier days of quantum mechanics there has been continual
interest in the models for which the corresponding Schr\"odinger 
equation can be solved exactly. The  number of totaly exactly
solvable potentials is rather limited. Therefore, recently much 
attention has been given to the quasi-exactly solvable (QES) potentials
for which a finite number of the energy levels and the corresponding
wave functions are known in the explicit form.

The first examples of QES potentials were given in 
\cite{Sig78,Fle81,Raz80,Kha81}.
Subsequently several methods were worked out for generating QES potentials
and as a result many QES potentials were established 
\cite{Tur87,Tur88Zh,Tur88,Shi89,Ush94,Zas83,Zas84,Uly97,Gan95}.
One of the methods is the generation of new QES potentials using 
supersymmetric (SUSY) quantum mechanics 
\cite{Gan95,Jat89,Roy91}
(for review of SUSY quantum mechanics see \cite{Coo95}).
The idea of the SUSY method for constructing QES potentials is the following. 
Starting from some initial QES potential with $n+1$ known eigenstates and 
using the properties of the unbroken SUSY one obtains the supersymmetric 
partner potential which is a new QES 
potential with the $n$ known eigenstates.

In our recent paper \cite{Tka98} we have 
proposed a new SUSY method for generating 
QES potentials with two known eigenstates. This method, in 
contrast to the one in papers \cite{Gan95,Jat89,Roy91}, does not require 
the knowledge of the initial QES potential for generating a new QES one. 
The general expression for the superpotential, the potential energy
and two wave functions which correspond to two energy levels
were obtained. Within the frames of this method we have obtained
QES potentials for which we have found in the explicit form the energy
levels and wave functions of the ground and first excited states. 
One should mention here also
paper \cite{Cat95} where the general expression for the 
QES potentials with two known eigenstates was obtained without resorting
to the SUSY quantum mechanics 
(see also a very recent paper by Dolya and Zaslavskii \cite{Dol01}). 
Although this method is direct and 
simpler than the SUSY approach the latter still has some advantages.
Namely, the SUSY method developed in \cite{Tka98} can be extended for the 
generation of QES potentials with three known eigenstates \cite{Kul99}
and conditionally exactly solvable potentials (CES) \cite{Tka99Ph}.
The CES potentials are those for which the eigenvalue problem
for the corresponding Hamiltonian is exactly solvable only when the potential
parameters obey certain conditions \cite{Sou93}.
About using SUSY quantum mechanics for the construction
of the CES potentials see \cite{Nag94,Jun97,Jun98}.
It is worth to mention the very recent paper \cite{Bri01} where the authors
established the connection between SUSY approach for constructing
QES potentials with two known eigenstates \cite{Tka98,Tka99Ph} and
the Turbiner approach \cite{Tur87}.  

Note that the general expressions for QES potential and corresponding 
wave functions derived by Dolya and Zaslavskii in \cite{Dol01} 
without resorting to SUSY quantum mechanics
are the same
as those obtained in \cite{Tka98,Tka99Ph} using SUSY method.
In \cite{Tka98,Tka99Ph} we have used these general expression for
constructing QES potentials with the ground and first excited states. 
A new interesting result obtained by Dolya and Zaslavskii is
that they have shown that it is possible to obtain
not only the ground and first excited states but
any pair of the energy levels and the corresponding wave functions.

The aim of the present paper is to extend the SUSY method proposed in
our papers \cite{Tka98,Tka99Ph} for 
constructing QES potentials 
with arbitrary two known eigenstates.
In \cite{Tka98,Tka99Ph} we used nonsingular superpotentials and
obtained QES potentials with explicitly known ground and first excited
states. In the present paper using singular superpotentials we obtain
nonsingular QES potentials for which we know in the explicit form
any pair of the energy levels and the corresponding wave functions.
Of course, the idea 
of using singular superpotential in SUSY quantum mechanics
is well know (see \cite{Coo95} for review).
Nevertheless, 
a new moment of present paper is that in the frame of SUSY quantum mechanics 
with singular superpotentials we derive nonsingular QES potentials
with arbitrary two known eigenstates. 
 
\section{SUSY quantum mechanics and QES problems}

In the Witten's model of supersymmetric quantum mechanics
the SUSY partner Hamiltonians $H_\pm$ read
\begin{equation} \label{SUSYH}
H_\pm=B^\mp B^\pm=-{1\over2}{d^2\over dx^2}+ V_\pm(x),
\end{equation}
where
\begin{eqnarray} \label{Bpm}
B^\pm={1\over\sqrt{2}}\left(\mp{d\over dx}+ W(x)\right), \\ \label{SUSYV}	
V_\pm (x)={1\over 2}\left(W^2(x) \pm W'(x)\right), \ \ W'(x)={dW(x)\over
dx},
\end{eqnarray}
and $W(x)$ is referred to as a superpotential.
In this paper we shall consider the  systems 
on the full real line 
$-\infty<x<\infty$.

We shall study the eigenvalue problem for the Hamiltonian $H_-$
\begin{equation} \label{SroHm}
B^+B^-\psi^-_E(x)=E\psi^-_E(x).
\end{equation}
Applying to the left and right hand sides of this equation 
the operator $B^-$ we obtain the equation
for the eigenvalue problem of the Hamiltonian $H_+$
\begin{equation} \label{SroHp}
B^-B^+(B^-\psi^-_E(x))=E(B^-\psi^-_E(x))
\end{equation}
from which follows that 
\begin{equation} \label{trpm}
\psi^+_E(x)=CB^-\psi^-_E(x)
\end{equation}
is the solution of the eigenvalue problem for the Hamiltonian $H_+$
with the energy $E$, $C=1/\sqrt E$ is the normalisation constant
if the wave function $\psi^+(x)$ is square integrable.
It is also possible that $\psi^+(x)$ does not 
satisfy the necessary conditions
and does not belong to the eigenfunctions
of the Hamiltonian $H_+$.
Such situation takes place for singular superpotentials 
(see review \cite{Coo95}).
Nevertheless, the function $\psi^+(x)$ is the solution 
of the equation (\ref{SroHp}).
Applying to equation (\ref{trpm})
the operator $B^+$ we obtain
\begin{equation} \label{trmp}
\psi^-_E(x)=CB^+\psi^+_E(x).
\end{equation}

Now let us analyse the eigenvalue problem for the Hamiltonian $H_-$.
Due to the factorization of the Hamiltonian the wave function of 
the zero energy state satisfies the equation $B^-\psi^-_0=0$ and reads
\begin{equation} \label{psi0}
\psi_0^-(x)=C^-_0\ \exp\left(-\int W(x) dx\right),
\end{equation}  
$C^-_0$ is the normalization constant.
In order to satisfy
the condition of the square integrability of the wave function 
(\ref{psi0}) we put
\begin{equation} \label{signW}
{\rm sign}(W(\pm \infty)) = \pm 1.
\end{equation}

We are interested in the potential energy $V_-(x)$ free of singularities.
The simplest way to satisfy this condition is to consider a superpotential
$W(x)$ which is free of singularities. Then $\psi^-_0(x)$ corresponds 
to the zero energy ground state of the Hamiltonian $H_-$.
Just this case was considered in our paper \cite{Tka98}.
But it is also possible to get a nonsingular potential 
energy $V_-(x)$ using a singular superpotential. 
Let us assume that $W(x)$ has 
the simple poles at the points $x_k$ with the 
following behaviour in the vicinity of $x_k$
\begin{equation} \label{Was}
W(x)={A_{-1}\over x-x_k} +A_0 +A_{1}(x-x_k)+ {\it O}((x-x_k)^2).
\end{equation}
Then $V_-(x)$ in the vicinity of $x_k$ reads
\begin{equation}
2V_-(x)={A_{-1}(A_{-1}+1)\over (x-x_k)^2}+2{A_{-1}A_0\over x-x_k}+ 
2A_{-1}A_1+A_0^2-A_1+ {\it O}(x-x_k).
\end{equation}
The first case $A_{-1}=0$ leads to the nonsingular $W(x)$ and $V_-(x)$.
The second case $A_{-1}=-1$ and $A_0=0$ gives the
nonsingular potential energy $V_-(x)$ with a singular superpotential.
Here it is worth stressing that the SUSY partner potential energy
in this case is singular with the following behaviour in 
the vicinity of $x_k$: $2V_+(x)=2/(x-x_k)^2-A_1 +O(x-x_k)$.

Let us analyse the second case.
Using equation (\ref{psi0}) we obtain the behaviour of the wave function
in the vicinity of the point $x_k$ as follows 
$\psi^-_0(x)\sim |x-x_k|(1-A_1(x-x_k)^2/2)$.
As we see $d\psi^-_0(x)/dx$ is a discontinuous function at the points $x_k$.
In order to obtain the wave function with continuous derivative we use
the simple fact that if $\psi^-_0(x)$ in the domain $x_k<x<x_{k+1}$
satisfies the Schr\"odinger equation then 
the function with the opposite sign $-\psi^-_0(x)$ satisfies the
same equation too. 
Thus, we can change the sign of wave function
in the domains $(x_k,x_{k+1})$ in such a way that
$\psi^-_0(x)$ and its derivative $d\psi^-_0(x)/dx$ will be
continuous functions. 
In fact it means that it is necessary to 
make the substitution $|f|\to f$.
Then the behaviour of the wave function in 
the vicinity of $x_k$ is
\begin{equation} \label{aspsi}
\psi^-_0(x)\sim (x-x_k)(1-A_1(x-x_k)^2/2)
\end{equation}
and the wave function has zeros at the points $x_k$.
Thus, the zero energy wave function has $n$ nodes
($n$ is the number of poles of the superpotential) 
and corresponds to the $n$-th excited state.
Note that in this case the ground state energy is less than zero.
Choosing different superpotentials $W(x)$ we can easily construct 
different QES potentials with one known eigenstate.

In contrast to this the construction of the QES potentials with
two known eigenstates is not a trivial problem. One state of 
the Hamiltonian $H_-$ with the zero energy
is known and is given by (\ref{psi0}).
In order to obtain one more state of $H_-$ 
we use the following
well-known procedure exploited in SUSY quantum mechanics. 
Let us consider the SUSY partner of $H_-$, i.e. the Hamiltonian $H_+$. 
If we calculate some state of $H_+$ we immediately find 
a new excited state of $H_-$ using 
transformation (\ref{trmp}). 
In order to calculate 
some state of $H_+$ let us rewrite it in the following form
\begin{equation} \label{Hpm}
H_+=H_-^{(1)} + \epsilon = B_1^+B_1^- + \epsilon, \ \  \epsilon > 0,
\end {equation}
which leads to the following relation between the potential energies
\begin{equation}\label{Vpm}
V_+(x)= V^{(1)}_-(x)+\epsilon,
\end{equation}
and superpotentials
\begin{equation} \label{WW}
W^2(x)+W'(x)=W_1^2(x)-W'_1(x) +2 \epsilon,
\end{equation}
where 
$\epsilon$ is the energy of the state of $H_+$ 
since we supposed that $H_-^{(1)}$ 
similarly to $H_-$
has the zero energy state,
$B_1^{\pm}$ and $V^{(1)}_-(x)$ are given by 
(\ref{Bpm}) and (\ref{SUSYV})
with the new superpotential $W_1(x)$.  

As we see from (\ref{Hpm}) the wave function 
of $H_+$ with the energy $E=\epsilon$ is also 
the zero-energy wave function of $H_-^{(1)}$ 
and it satisfies the equation
$B^-_1\psi_\epsilon^+(x)=0$.
The solution of this equation is
\begin{equation} \label{psip}
\psi^+_{\epsilon}(x)=C^+ \ \exp\left(-\int W_1(x) dx\right).
\end{equation}  
Using (\ref{trmp}) we obtain 
the wave function of the excited state with the energy level
$E=\epsilon$ for the Hamiltonian $H_-$
\begin{equation} \label{psi1}
\psi_{\epsilon}^-(x)=C^-\ W_+(x) \exp\left(-\int W_1(x) dx\right)
\end{equation}  
where we have introduced the notation
$W_+(x)=W_1(x)+W(x)$.
In order to satisfy the square integrability of this function
at infinity we impose the superpotential $W_1(x)$ 
with the same condition as $W(x)$ (\ref{signW}). 
Then $W_+(x)$ satisfies the same condition (\ref{signW}) too.

In order to obtain the explicit expression for the wave function  
$\psi^-_0(x)$ with zero energy
and the wave function $\psi^-_{\epsilon}(x)$
with the energy $\epsilon$ given by (\ref{psi0}) and (\ref{psi1})
it is necessary to obtain the explicit expression for superpotentials
$W(x)$ and $W_1(x)$.
This is the subject of the next section.

\section{Solutions for superpotentials and 
construction of nonsingular QES potentials}
The superpotentials $W(x)$ and $W_1(x)$ satisfy equation (\ref{WW}).
Note, that (\ref{WW}) is the Riccati equation 
which cannot be solved exactly
with respect to $W_1(x)$ for a given $W(x)$
and vice versa. 
But we can find such a pair of $W(x)$ and
$W_1(x)$ that satisfies equation (\ref{WW}). For this purpose
let us rewrite equation (\ref{WW}) in the following form
\begin{equation} \label{EqWpWm}
W'_+(x)=W_-(x)W_+(x) +2\epsilon,
\end{equation}
where
\begin{eqnarray} \label{Wp}
W_+(x)=W_1(x) + W(x),\\ \label{Wm}
W_-(x)=W_1(x) - W(x). 
\end{eqnarray}

This new equation can be easily solved with respect to $W_-(x)$
for a given $W_+(x)$ and vice versa.
In this paper we use the solution of equation (\ref{EqWpWm}) with
respect to $W_-(x)$
\begin{equation} \label{SolWm}
W_-(x)=(W'_+(x) - 2\epsilon )/W_+(x).
\end{equation}
Then from (\ref{Wm}), (\ref{Wp}) and (\ref{SolWm}) we obtain the pair of
$W(x)$, $W_1(x)$ that satisfies equation (\ref{WW})
\begin{eqnarray}  \label{SolW0}
W(x)={1\over 2}\left(W_+(x) - (W'_+(x)-2\epsilon)/W_+(x) \right), \\  
\label{SolW1}
W_1(x)={1\over 2}\left(W_+(x) + (W'_+(x)-2\epsilon)/W_+(x) \right), 
\end{eqnarray}
here $W_+(x)$ is some function of $x$ which generates the superpotentials
$W(x)$ and $W_1(x)$.
Let us stress that $W(x)$, $W_1(x)$ and $W_+(x)$ must satisfy condition
(\ref{signW}).
It is necessary to note that
the general solutions (\ref{SolW0}), (\ref{SolW1}) 
of equation (\ref{WW})
was obtained earlier in \cite{Bec93} in 
the context of parasupersymmetric quantum mechanics.

In our earlier paper \cite{Tka98} we considered only free of 
the singularities superpotentials $W(x)$ and $W_1(x)$.
To satisfy a nonsingularity of the superpotentials
we considered a continuous function
$W_+(x)$ that has only one simple zero. 
Because $W_+(x)$ was considered as continuous function which satisfies 
condition (\ref{signW}) 
the function $W_+(x)$ must have at least one zero. 
Then, as we see from (\ref{SolWm}), (\ref{SolW0}) and (\ref{SolW1}),
$W_-(x)$, $W(x)$ and $W_1(x)$ have the poles. 
In order to construct the superpotentials free of singularities
we supposed that $W_+(x)$ has only one simple zero at $x=x_0$.
In this case the pole of $W_-(x)$ and $W(x)$, $W_1(x)$ at $x=x_0$ can
be cancelled by choosing
$\epsilon = W'_+(x_0)/2$.

Now let us consider more general cases of the 
function $W_+(x)$ which leads 
to nonsingular QES potential energy $V_-(x)$.

{\it Case 1}.
Suppose that $W_+(x)$ has simple zeros at the points $x_k$, 
$k=1,...,n$
\begin{equation} \label{Wplus}
W_+(x)=W'_+(x_k)(x-x_k)+{1\over 2}W''_+(x_k)(x-x_k)^2+ {\it O}((x-x_k)^3).
\end{equation}
The zeros of the function $W_+(x)$ lead to the poles of 
the function $W_-(x)$
and the superpotential
$W(x)$ with the following behaviour in the vicinity of $x_k$
\begin{equation} \label{Was1}
W(x)=
-\left({1\over2}-{\epsilon\over W'_+(x_k)}\right){1\over x-x_k}-
{1\over 2}{W''(x_k)\over W'(x_k)}
\left({1\over2}+{\epsilon\over W'_+(x_k)}\right)
+{\it O}(x-x_k).
\end{equation}
The behaviour of the superpotential $W_1(x)$ 
in the vicinity of $x_k$ is similar to $W(x)$ only with the 
opposite sign.
It is worth comparing the superpotential (\ref{Was1}) with (\ref{Was}):
$A_{-1}={\epsilon/ W'_+(x_k)}-{1/2}$ and 
$A_0=-W''(x_k)({\epsilon/ W'_+(x_k)}+{1/2})/2W'(x_k)$.
This superpotential leads to the following behaviour of the potential
energy in the vicinity of $x_k$
\begin{equation}
2V_-(x)=\left[\left({\epsilon\over W'_+(x_k)}\right)^2
-{1\over4}\right]\left({1\over (x-x_k)^2}
-{W''(x_k)\over W'(x_k)}{1\over(x-x_k)}\right)
+{\it O}({\rm const}).
\end{equation}
Thus in the case 
\begin{equation} \label{Wst}
W'_+(x_k)=\pm 2\epsilon
\end{equation}
the potential energy $V_-(x)$ is free of singularities.
It is convenient to divide the set of $x_k$ into two subsets
$x^+_k$ $(k=1,...,n^+)$ and 
$x^-_k$ $(k=1,...,n^-)$ for which 
$W'_+(x^+_k)=2\epsilon >0$ and
$W'_+(x^-_k)=-2\epsilon <0$. 
We suppose in this paper that $\epsilon >0$. 
Because of $W'_+(x^+_k)=2\epsilon$ the singularity at the points 
$x^+_k$ is cancelled and $W(x)$, $W_1(x)$ have singularities 
only at the points $x^-_k$
\begin{eqnarray}
W(x)={-1\over x-x^-_k} + O(x-x^-_k), \\
W_1(x)={1\over x-x^-_k} + O(x-x^-_k). \label{W1pol}
\end{eqnarray}
Substituting $W(x)$ into (\ref{psi0}) and
using the result of the previous section (see Eq. (\ref{aspsi})) 
we see that the wave function
$\psi^-_0(x)$ with zero energy has $n^-$ zeros at the points $x^-_k$,
namely $\psi^-_0 \sim (x-x^-_k)$ in the vicinity of $x^-_k$.
Substituting $W_+(x)$ given by (\ref{Wplus})
and $W_1(x)$ given by (\ref{W1pol}) into (\ref{psi1})
we obtain that the wave function $\psi^-_\epsilon(x)$ 
with the energy $\epsilon$ 
has $n^+$ zeros at the points $x^+_k$: 
$\psi^-_\epsilon(x)\sim (x-x^+_k)$.
When $W_+(x)$ is the continuous function satisfying condition 
(\ref{signW}) then $n^+=n^-+1$.
Thus in this case $\psi^-_0(x)$ and $\psi^-_{\epsilon}(x)$ correspond to
$n^-$-th and $(n^-+1)$-th excited states respectively.

{\it Case 2}.
Now let us assume that
the function $W_+(x)$ in addition to the zeros 
has the simple poles at the points $x^0_k$ 
with the behaviour in the vicinity of $x^0_k$ as follows
\begin{equation}
W_+(x)={G_{-1}\over x-x^0_k}+G_0+{\it O}(x-x^0_k).
\end{equation}
Then
\begin{eqnarray}
W(x)={1\over 2}{G_{-1}+1\over x-x^0_k}+ \label{C2W}
{1\over 2}{G_0\over G_{-1}}(G_{-1}-1)+{\it O}(x-x^0_k), \\
W_1(x)={1\over 2}{G_{-1}-1\over x-x^0_k}+
{1\over 2}{G_0\over G_{-1}}(G_{-1}+1)+{\it O}(x-x^0_k),
\end{eqnarray}
here we drop out the terms of the order $(x-x^0_k)$.
Note that $G_0$ and $G_{-1}$ can depend on $k$.
For the sake of simplicity we are omitting this dependence.
Comparing superpotential (\ref{C2W}) and (\ref{Was}) we conclude
that the superpotential
$W(x)$ (\ref{C2W}) gives a nonsingular potential energy
for the {\it case 2a}: $G_{-1}=-1$ and $G_0$ is an arbitrary constant,
and for the {\it case 2b}: $G_{-1}=-3$ and $G_0=0$.

{\it Case 2a}.
For the case $G_{-1}=-1$ in the vicinity of $x^0_k$
we have nonsingular superpotential $W(x)$
and singular $W_1(x)$ 
\begin{eqnarray}
W(x)=G_0+O(x-x^0_k), \\
W_1(x)={-1\over x-x^0_k} +O(x-x^0_k).
\end{eqnarray}
The wave functions $\psi^-_0(x)$ and 
$\psi^-_\epsilon (x)$ calculated with these superpotentials
do not have the zeros at the points $x^0_k$,
which we will denote in this case as $a_k$, $k=1,...,n^0$.
Nevertheless, the important moment is that now $W_+(x)$ 
in additional to $n=n^+ +n^-$ zeros at points $x^+_k$ and $x^-_k$
has $n^0$ poles at the points $x^0_k$
and thus is not a continuous function. 
As a result in this case we have
$n^+=n^-+n^0+1$.
Using the result obtained in the case 1 we see that
$\psi^-_0(x)$ has the zeros at $x^-_k$
and corresponds to the $n^-$-th excited state and
$\psi^-_\epsilon$ has the zeros at $x^+_k$
and corresponds to the $(n^-+n^0+1)$-th excited state.

{\it Case 2b}.
The case $G_{-1}=-3$ and $G_0=0$ leads to 
the following behaviour of superpotentials in the vicinity of $x^0_k$
\begin{eqnarray}
W(x)={-1 \over x-x^0_k} +O(x-x^0_k), \\
W_1(x)={-2 \over x-x^0_k} +O(x-x^0_k)
\end{eqnarray}
The wave functions
$\psi^-_0(x)$ and $\psi^-_\epsilon(x)$ calculated with 
these superpotentials have common zeros at the points 
$x^0_k$, which we will denote in this case as $b_k$,
$k=1,...,m^0$.
Thus, when in addition to poles the function $W_+(x)$ has
$n=n^+ +n^-$ zeros at the the points $x^+_k$ and $x^-_k$
the wave function $\psi^-_0(x)$ corresponds to $(n^-+m^0)$-th
excited state and $\psi^-_\epsilon(x)$ 
corresponds to $(n^-+2m^0+1)$-th excited state.

Let us consider the {\it general case} 
which combines the cases 1, 2a and 2b. 
The function $W_+(x)$ has
$n^-$ zeros with negative derivatives at the points 
$x^-_k$ ($k=1,...,n^-$),
$n^0$ poles at the points $a_k$ ($k=1,...,n^0$)
with asymptotic behaviour in the vicinity
of these points $-1/(x-a_k)+{\rm const}$ 
and $m^0$ poles at the points $b_k$ ($k=1,...,m^0$)
with the asymptotic behaviour $-3/(x-b_k)$ .
The number of zeros of the function $W_+(x)$ with a positive derivative
at the points $x^+_k$
is the following $n^+=n^- +n^0 + m^0+ 1$.
The wave function $\psi^-_0(x)$
has the $(n^-+m^0)$ nodes at the points $x^-_k$ and $b_k$
and thus it corresponds to the $(n^-+m^0)$-th
excited state.
The wave function $\psi^-_\epsilon(x)$ 
has $n^+ +m^0=n^-+n^0+2m^0+1$ nodes at the points 
$x^+_k$ and $b_k$
and thus it corresponds to the $(n^-+n^0+2m^0+1)$-th excited state.

Thus, the considered cases 1, 2a, 2b and the combined general case  
lead to the nonsingular QES potential energy
$V_-(x)$ given by (\ref{SUSYV}),
where the superpotential $W(x)$ is expressed over the function $W_+(x)$
by equation (\ref{SolW0}).
The zero energy wave function $\psi^-_0(x)$
and the wave function $\psi^-_\epsilon(x)$ 
with the energy $\epsilon$   
are given by (\ref{psi0}) and (\ref{psi1}), respectively.
Note, that in the case of nonsingular superpotential $W(x)$ the zero energy
wave function corresponds to the ground state,
but in the case of singular superpotential 
the energy of the ground state is less than zero 
and the zero energy wave function corresponds to an excited state.

In the conclusion of this section let us discuss the second
possibility for construction of
the QES potentials with two known eigenstates.
Namely, we can use the solution of equation (\ref{EqWpWm})
with respect to $W_+(x)$ \cite{Tka99Ph}.
We found that 
\begin{equation} \label{Wphi}
W_-(x)=-{\phi''(x)\over\phi'(x)}, \ \ 
W_+(x)=2\epsilon {\phi(x)\over\phi'(x)}
\end{equation}
satisfies equation (\ref{EqWpWm}), where $\phi(x)$ is new
generating function.
Using $W_+$ and $W_-$ given by (\ref{Wphi}) we obtained the QES
potential and two wave functions in term of $\phi(x)$.
Choosing generating functions $\phi(x)$ 
with one zero we obtained QES potentials
with explicitly know ground and first excited states.
Note that basic equation derived by Dolya and Zaslavskii in 
\cite{Dol01} without resorting to SUSY quantum mechanics 
are the same as was 
earlier obtained in \cite{Tka99Ph} using SUSY method
(in \cite{Dol01} $\phi(x)$ is denoted as $\xi(x))$.
A new result obtained by Dolya and Zaslavskii is
that they have shown how one can obtain not only the ground and
first excited states but any pair of states
using generating function $\phi(x)$ and their derivative $\phi'(x)$
with zeros and poles. 
In the present paper we are working with generating function $W_+(x)$
which is related with $\phi(x)$ by (\ref{Wphi}).
As we see zeros and poles of $\phi(x)$ (or $\xi(x)$)
lead the zeros of $W_+(x)$ and zeros of $\phi'(x)$ (or $\xi'(x)$)
lead to the poles of $W_+(x)$. 

\section{Examples of QES potentials}

Note, that  
all expressions depend on the function $W_+(x)$. 
We may choose various functions $W_+(x)$ and obtain as a result
various QES potentials. 
Note also that when the function $W_+(x)$ generates the potential energy
$V_-(x)$ then $W_+(x/a)/a$ generates the potential energy
$V_-(x/a)/a^2$. This scaling is useful for comparing,
in principle, the same potential
energies but written in different forms as a result of a different
measurement units.  

To illustrate the above described method we give two explicit examples
of the nonsingular QES potentials.

{\it Example 1}

Let us consider a continuous function $W_+(x)$ which corresponds 
to case 1
\begin{equation}
W_+(x)=\alpha x{x^2-1\over x^2+1},
\end{equation}
where $\alpha >0$. 
This function has three zeros at the points $0, \pm 1$.
The denominator is written in order to satisfy condition
(\ref{Wst}), namely $W'_+(0)=-\alpha$, $W'_+(\pm 1)=\alpha$.
Note that $n^+=2$, $n^-=1$
and thus we have QES potential with explicitly known
first and second excited states.
From the condition of nonsingularity
of the potential energy it follows that
$\epsilon=\alpha/2$. 
Then using (\ref{SolW0}) and (\ref{SolW1}) we obtain 
for the superpotentials
\begin{eqnarray}
W(x)={\alpha \over 2}x+(1-\alpha){x\over x^2+1}-{1\over x}, \\
W_1(x)={\alpha \over 2}x-(1+\alpha){x\over x^2+1}+{1\over x}.
\end{eqnarray}
The superpotential $W(x)$ gives the following QES potential
\begin{equation} \label{Ex1V}
2V_-(x)={\alpha^2\over 4}x^2
-(1-\alpha)(3-\alpha){1\over(x^2+1)^2}
-2\alpha(1-\alpha){1\over x^2+1} 
+\alpha(1-\alpha)- {3\over 2}\alpha.
\end{equation}

The zero energy wave function (\ref{psi0}) and wave function with energy 
$\epsilon =\alpha /2$ (\ref{psi1}) read
\begin{eqnarray}
\psi^-_0(x)=
C_0 x(x^2+1)^{(\alpha-1)/2} \exp \left(-{\alpha\over 4} x^2\right),\\
\psi^-_\epsilon(x)=C_\epsilon 
(x^2-1)(x^2+1)^{(\alpha-1)/2} \exp \left(-{\alpha\over 4} x^2\right).
\end{eqnarray}
As we see $\psi^-_0(x)$ has one node and thus really corresponds
to the first excited state, $\psi^-_\epsilon (x)$ has two nodes and
corresponds to the second excited state.

Note that QES potential (\ref{Ex1V}) has similar structure as the
potential studied in \cite{Dol01} but in fact it is another potential.
In our case we have the QES potential with explicitly known
first and second eigenstates, whereas for the QES potential
studied in \cite{Dol01} the ground 
and second excited states are explicitly known.
It is interesting to note also that in our case 
QES potential (\ref{Ex1V}) at the value of
parameter $\alpha=1$ becomes exactly solvable and
corresponds to the harmonic oscillator.

{\it Example 2}

Let us consider a more complicated example for which the function
$W_+(x)$ has two poles and three zeros
\begin{equation}
W_+(x)={\alpha \over x^2-1}x(x^2-a^2)(x^2+b^2).
\end{equation}   
In order to have the asymptotic behaviour 
of the function $W_+(x)$ in the vicinity
of the points $x=\pm 1$ which corresponds to the case 2a,
namely $-1/(x-1)$ and $-1/(x+1)$,
we choose the parameter $\alpha$ as follows 
\begin{equation}
\alpha={2\over (a^2-1)(b^2+1)}.
\end{equation}
This value of $\alpha$ gives a nonsingular behaviour of the potential energy
in the vicinity of $x=\pm 1$.
In order to have the same derivatives in the points of zeros of 
the function $W_+(x)$, namely $W'_+(0)=W'_+(\pm a)$,
we put the following value for parameter $b$
\begin{equation}
b^2={2a^2\over(a^2-3)},
\end{equation}
from which it follows that $a^2 > 3$.
Then choosing
\begin{equation}
2\epsilon = W'_+(0)= W'_+(\pm a)= {4 a^4 \over 3(a^2-1)^2}
\end{equation}
we obtain the superpotentials
\begin{eqnarray}
W(x)={x\over 6}\left(-{2(a^4-6a^2+3)\over (a^2-1)^2}
+{2(a^2-3)x^2\over (a^2-1)^2}
-{15(a^2-3)\over (a^2-3)x^2+2a^2}\right)  \\
W_1(x)={x\over 6}\left(-{2(a^4-6a^2+3)\over (a^2-1)^2}
+{2(a^2-3)x^2\over (a^2-1)^2}
+{15(a^2-3)\over (a^2-3)x^2+2a^2}-{12\over x^2-1}\right)
\end{eqnarray}
and the nonsingular potential energy
\begin{eqnarray}
2V_-(x)=
{x^2\over 36}\left(-{2(a^4-6a^2+3)\over (a^2-1)^2}
+{2(a^2-3)x^2\over (a^2-1)^2}
-{15(a^2-3)\over (a^2-3)x^2+2a^2}\right)^2 \\ \nonumber
+{1\over 6}\left( 
{2(a^4-6a^2+3)\over (a^2-1)^2}-
{6(a^2-3)x^2\over (a^2-1)^2}+
{15(a^2-3)\over (a^2-3)x^2+2a^2}-
{30(a^2-3)^2x^2\over ((a^2-3)x^2+2a^2)^2}
\right).
\end{eqnarray}
For this QES potential we know explicitly 
the wave functions of the ground and the third excited states
\begin{eqnarray}
\psi^-_0(x)=C_0((a^2-3)x^2+2a^2)^{5/4}
\exp\left( {-(a^2-3)x^4
+2(a^4-6a^2+3)x^2\over 12(a^2-1)^2}\right), \\
\psi^-_\epsilon (x)=C_\epsilon 
{x(x^2-a^2)\over ((a^2-3)x^2+2a^2)^{1/4}}
\exp\left( {-(a^2-3)x^4
+2(a^4-6a^2+3)x^2\over 12(a^2-1)^2}\right).
\end{eqnarray} 
Note that at $a^2=3$ this potential becomes exactly solvable and
corresponds to the harmonic oscillator.

\section{Conclusions}

We propose the SUSY method for constructing the QES potentials
with arbitrary two known energy levels and corresponding wave functions.
This is an extension of our SUSY method proposed in \cite{Tka98,Tka99Ph}
where QES potentials with the ground and first excited states were
obtained.
In the proposed method the function $W_+(x)$ plays the role of 
a generating function. 
Choosing different functions $W_+(x)$ we obtain different
QES potentials $V_-(x)$. 
The two known wave functions $\psi^-_0(x)$
and $\psi^-_\epsilon (x)$ correspond to the eigenstates with zero energy
and  the energy $\epsilon$, respectively.
The state numbers of these wave functions depend on the properties
of function $W_+(x)$ as described in the cases 1, 2a, 2b which is
summarized in the general case in section 3.
In section 4 we consider the explicit examples of QES potentials with 
the rational generating function $W_+(x)$.
These examples yield some new one-parametric QES potentials. 
At some special values of parameters these potentials become 
exactly solvable and 
correspond to the harmonic oscillator.

We can consider various new generating functions 
$W_+(x)$ and obtain new QES potentials.
One of the interesting possibilities is to consider
the periodic functions $W_+(x)$ which lead to the periodic QES potentials.
This problem will be the subject of a separate paper.



\begin{thebibliography}{18}
\bibitem{Sig78} V.~Singh, S.~N.~Biswas, K.~Dutta, Phys. Rev. D {\bf 18} 
            (1978) 1901.
\bibitem{Fle81} G.~P.~Flessas, 
          Phys. Lett. A {\bf 72} (1979) 289.
\bibitem{Raz80} M.~Razavy, Am. J. Phys. {\bf 48} (1980) 285; 
          Phys. Lett A {\bf 82} (1981) 7.
\bibitem{Kha81} A.~Khare, Phys. Lett. A {\bf 83} (1981) 237.
\bibitem{Tur87} A.~V.~Turbiner, A.~G.~Ushveridze, Phys. Lett. A {\bf 126} 
             (1987) 181.
\bibitem{Tur88Zh} A. V. Turbiner,
               Zh. Eksp. Teor. Fiz. {\bf 94} (1988) 33.
\bibitem{Tur88} A.~V.~Turbiner, Commun. Math. Phys. {\bf 118} (1988) 467.
\bibitem{Shi89} M.~A.~Shifman, Int. Jour. Mod. Phys. A {\bf 4} (1989) 2897.
\bibitem{Ush94} A.~G.~Ushveridze, Quasi-exactly solvable models in quantum 
            mechanics,
            Institute of Physics Publishing, Bristol (1994).
\bibitem{Zas83} O.~B.~Zaslavskii, V.~V.~Ul'yanov, V.~M.~Tsukernik,
             Fiz. Nizk. Temp. {\bf 9} (1983) 511.
\bibitem{Zas84} O.~B.~Zaslavsky, V.~V.~Ulyanov, Zh. Eksp. Teor. Fiz.
             {\bf 87} (1984) 1724.
\bibitem{Uly97} V.~V.~Ulyanov, O.~B.~Zaslavskii, J.~V.~Vasilevskaya,
            Fiz. Nizk. Temp. {\bf 23} (1997) 110. 
\bibitem{Gan95} A.~Gangopadhyaya, A.~Khare, U.~P.~Sukhatme,
             Phys. Lett. A {\bf 208} (1995) 261.
\bibitem{Jat89} D.~P.~Jatkar, C.~Nagaraja Kumar, A.~Khare, Phys. Lett. A 
             {\bf 142} (1989) 200.
\bibitem{Roy91} P.~Roy, Y.~P.~Varshni, Mod. Phys. Lett. A {\bf 6}
             (1991) 1257.
\bibitem{Coo95} F.~Cooper, A.~Khare, U.~Sukhatme, Phys. Rep. {\bf 251} 
             (1995) 267.
\bibitem{Tka98} V.~M.~Tkachuk, Phys. Lett. A {\bf 245} (1998) 177.
\bibitem{Cat95} A. Caticha, Phys. Rev.A {\bf 51} (1995) 4264.
\bibitem{Dol01} S. N. Dolya, O. V. Zaslavskii, 
             J. Phys. A {\bf 34} (2001) 1981.
\bibitem{Kul99} T. V. Kuliy, V. M. Tkachuk, J. Phys. A {\bf 32}
             (1999) 2157.
\bibitem{Tka99Ph} V. M. Tkachuk, J. Phys. A {\bf 32} (1999) 1291.

\bibitem{Sou93} A.~de Souza Dutra, Phys. Rev. A {\bf 47} (1993) R2435.
\bibitem{Nag94} N.~Nag, R.~Roychoudhury, Y.~P.~Varchni, Phys. Rev. A
              {\bf 49} (1994) 5098.
\bibitem{Jun97}  G. Junker, P. Roy, Phys. Lett. A {\bf 232} (1997) 113.
\bibitem{Jun98}  G. Junker, P. Roy, Ann. Phys., NY {\bf 270} (1998) 155.
\bibitem{Bri01} Y. Brihaye, N. Debergh, J. Ndimubandi, quant-ph/0104009.
\bibitem{Bec93} J.~Beckers, N.~Debergh, A.~G.~Nikitin, Mod. Phys. Lett. A 
{\bf 8} (1993) 435.

\end{thebibliography}
\end{document}